\documentclass{emulateapj}
\usepackage{apjfonts}
\usepackage{graphicx}

\def\apj{\rm{ApJ}}
\def\mnras{\rm{MNRAS}}%
\def\apjl{\rm{ApJ}}%
\def\aap{\rm{A\&A}}%
\def\beq{\begin{equation}} 
\def\eeq{\end{equation}}

\def\mfp{\lambda_{\rm abs}}
\def\deltat{\tilde{\delta}}

\shorttitle{ABSORPTION SYSTEMS \& REIONIZATION}
\shortauthors{ALVAREZ \& ABEL}

\begin{document}

\title{The Effect of Absorption Systems on Cosmic Reionization}

\author{Marcelo A. Alvarez\altaffilmark{1} and Tom Abel\altaffilmark{2}}

\altaffiltext{1}{Canadian Institute for Theoretical Astrophysics,
  University of Toronto, 60 St.George St., Toronto, ON M5S 3H8,
  Canada} 
\altaffiltext{2}{Kavli Institute for Particle Astrophysics and
  Cosmology, Stanford University, Menlo Park, CA 94025, USA}
\email{malvarez@cita.utoronto.ca}

\begin{abstract}
We use large-scale simulations to investigate the morphology of
reionization during the final, overlap phase. Our method uses an
efficient three-dimensional smoothing technique which takes into
account the finite mean free path due to absorption systems, $\mfp$,
by only smoothing over scales $R_s<\mfp$.
The large dynamic range of our calculations is necessary to resolve
the neutral patches left at the end of 
reionization within a representative volume; we find that simulation
volumes exceeding several hundred Mpc on a side are necessary in
order to properly model reionization when the neutral fraction is
$\simeq 0.01-0.3$.  Our results indicate a
strong dependence of percolation morphology on a large and uncertain
region of model parameter space. The single most important parameter
is the mean free path to absorption systems, which serve as opaque
barriers to ionizing radiation.  If these absorption systems were as
abundant as some realistic estimates indicate, the spatial structure
of the overlap phase is considerably more complex than previously predicted.
In view of the lack of constraints on the mean free path at the highest redshifts,
current theories that do not include absorption by Lyman-limit
systems, and in particular three-dimensional simulations, may
underestimate the abundance of neutral clouds at the end of
reionization. This affects predictions for the 21 cm signal associated
with reionization, interpretation of absorption features in quasar
spectra at $z \sim ~5-6$, the connection between reionization and
the local universe, and constraints on the patchiness and duration of
reionization from temperature fluctuations measured in the cosmic
microwave background arising from the kinetic Sunyaev-Zel'dovich
effect.  
\end{abstract}


\section{Introduction}

On scales relevant to the large-scale structure of
reionization -- hundreds of Mpc -- numerical calculations cannot
self-consistently track all the sources and sinks of ionizing
radiation. Resolving the IGM on the small scales relevant to QSO
absorption systems optically-thick at the Lyman edge, also known as
Lyman limit systems \citep[e.g.,][]{weymann/etal:1981, tytler:1982,
  mo/miralda-escude:1996}, is particularly difficult.  These
absorption systems were likely to have impeded the progress of
reionization \citep[e.g.,][]{miralda-escude/etal:2000,
  gnedin/fan:2006}.  

What are likely values for the comoving absorption system mean free
path at the end of reionization, $\mfp$?  \citet{prochaska/etal:2009}
determined the abundance of Lyman-limit absorption systems,  $dN/dz =
1.9 [(1+z)/4.7]^{5.2}$, over the redshift range $3.6<z<4.4$, which
translates into a comoving  mean free path of $\mfp=c/H(z)/(dN/dz)
\sim 415$ Mpc at $z=3.7$.  This dependence on redshift is steeper than
reported previously \citep[e.g.,][]{storrie-lombardi/etal:1994}, and
implies $\mfp\propto (1+z)^{-6.7}$.  However,
\citet{songaila/cowie:2010} extended the constraints out to $z\sim 6$,
finding a much shallower dependence, $\mfp\propto (1+z)^{-3.44}$, with
a measurement of $\mfp\simeq 34$~Mpc at $z\sim 5.7$. Comparison of the
absorption spectra and hydrodynamic simulations can also constrain
$\mfp$, with \citet{bolton/haehnelt:2007a} reporting $20 <\lambda_{\rm
abs}<50$ Mpc at $z\sim 6$.  The evolution of the mean free path at
$z<6$ leaves open a wide range of possible values during the
reionization epoch, from several to hundreds of Mpc.   Our goal is to
determine the sensitivity of the history and morphology of
reionization to this wide range of possible mean free paths.    

Direct simulation of all the processes involved in reionization is not
currently possible. Theoretical studies focus either on the small scale
hydrodynamics and radiative transfer
\cite[e.g.,][]{miralda-escude/etal:2000, gnedin:2000,
  haiman/etal:2001, shapiro/etal:2004, ciardi/etal:2006,
  pawlik/etal:2009, raicevic/theuns:2011, mcquinn/etal:2011}, or on
the large scale morphology of reionization
\cite[e.g.,][]{2002MNRAS.330L..53A, furlanetto/etal:2004,
  2006MNRAS.369.1625I, mcquinn/etal:2007, trac/cen:2007,
  shin/etal:2008, friedrich/etal:2011}.
On the largest scales, it has been somewhat surprising how well
simplified  ``semi-numerical'' approaches
\citep[e.g.,][]{zahn/etal:2007, mesinger/furlanetto:2007,
  thomas/etal:2009, choudhury/etal:2009, alvarez/etal:2009, santos/etal:2010,
  zahn/etal:2011a, mesinger/etal:2011a}
match much more computationally expensive radiative transfer
techniques \citep[e.g.,][]{1999ApJ...523...66A,2001NewA....6..437G,
  sokasian/etal:2001, 2002MNRAS.330L..53A, 2006MNRAS.369.1625I,
  mcquinn/etal:2007, trac/cen:2007}.
In order to efficiently survey the wide parameter space of $\mfp$ and
focus on the large scale morphology and overall progress of
reionization, we have chosen the simplified semi-numerical approach. 
Our implementation is similar to that of \citet{zahn/etal:2007}, with
the the addition of a treatment of photon consumption in absorption
systems. Our approach, in which we treat abundance of absorbers as an
input parameter, is complementary to that taken by
\citet{crociani/etal:2011}, where the semi-numerical approach was used
to {\em determine} the large scale distribution of absorbers, rather
than to model the effect of the absorbers on the progress of
reionization, as we do here. 

Our results are parametrized by different values of for the minimum
source halo mass, ionizing efficiency of collapsed matter, and the
absorption system mean free path.  We describe our simulations in \S2,
followed by the main results in \S 3, ending with a discussion in \S
4.  All calculations were done using
$(\Omega_m,\Omega_\Lambda,h,\sigma_8,n_s)=(0.27,0.73,0.72,0.8,0.96)$,
consistent with WMAP 7-year data \citep{komatsu/etal:2011}.  All
distances are comoving.     

\section{Simulations}

Table 1 shows our simulation parameters: $\zeta$, the ionizing
efficiency, $\mfp$, the comoving absorption system mean
free path, $t_{\rm ev}$, the photoevaporation time for the
evolving $\mfp$ models, and $M_{\rm min}$, the minimum halo mass of
galaxies. Also shown are the redshifts when the
global ionized fraction equals 0.5, 0.9, and 1 -- $z_{0.5}$,
$z_{0.9}$, and $z_{\rm ov}$, respectively, and the bubble mean
free path when the neutral fraction is 0.1, $\lambda_0\equiv
\lambda_{\rm b}(x_{\rm HI}=0.1)$.   The parameters were varied to give
a Thomson scattering optical depth
of($\tau_-,\tau_0,\tau_+$)$=$($0.06,0.09,0.12$), corresponding
to the 2-$\sigma$ constraint from {\em  WMAP}
\citep{komatsu/etal:2011}, with the He~I fraction tracking H~I, and
instantaneous He~II reionization at  $z=3$.  All simulations used
$4096^3$ cells in a 2 Gpc$/h$ box.   

\begin{table}[t]
\caption{Simulation parameters}
\centering
\begin{tabular}{l c c c c c c c c}
\hline\hline
$\tau_{\rm es}$&$M_{\rm min}$&$\mfp$&$t_{\rm ev}$&$\zeta$&$z_{1/2}$&$z_{0.9}$&$z_{\rm ov}$&$\lambda_0$\\
&$[M_\odot]$&[Mpc/h]&[Myr]&&&&&[Mpc]\\
\hline
0.06&$10^8$ & 8 & - &  20 & 8.1 & 5.9 & 4.3 & 87\\
0.06&$10^8$ & 32 & - & 16 & 8 & 6.3 & 5 & 181\\
0.06&$10^8$ & 256 & - & 14.5 & 8 & 6.8 & 6.2& 294 \\
0.06&$10^9$ & 8 & - & 78 & 8.2 & 6.6 & 5.6 & 98\\
0.06&$10^9$ & 32 & - & 60 & 8.1 & 6.8 & 6 & 210\\
0.06&$10^9$  & 256 & - & 40 & 8.1 & 7.24 & 6.9 & 375\\
\hline
0.09&$10^8$  & 8 & - & 137 & 11 & 9.2 & 8 & 94\\
0.09&$10^8$ & 32 & - & 110 & 10.9 & 9.5 & 8.5 & 201\\
0.09&$10^8$  & 256 & - & 98 & 10.9 & 9.9 & 9.5 & 345\\
0.09&$10^9$ & 8 & - & 1200 & 11.1 & 9.8 & 9.1 & 105\\
0.09&$10^9$ & 32 & - & 920 & 11 & 10 & 9.3 & 232\\
0.09&$10^9$ & 256 & - & 820 & 11 & 10.3 & 10 & 443\\
\hline 
0.12&$10^8$ & 8 & - & 1050 & 13.5 & 12.5 & 11.1 & 99\\ 
0.12&$10^8$ & 32 & - & 850 & 13.5 & 12.2 & 11.4 & 216\\ 
0.12&$10^8$ & 256 & - & 750 & 13.4 & 12.6 & 12.2 & 396\\ 
0.12&$10^9$ & 8 & - & 2.2e4 & 13.5 & 12.5 & 11.9 & 109\\
0.12&$10^9$ & 32 & - & 1.7e4 & 13.5 & 12.6 & 12.1 & 248\\
0.12&$10^9$ & 256 & - & 1.5e4 & 13.5 & 12.9 & 12.7 & 500\\
\hline
0.09\tablenotemark{a}&$10^8$  & 8 & - & 102 & 10.9 & 9.7 & 8.3 & -\\
0.09\tablenotemark{a}&$10^8$ & 32 & - & 94 & 10.8 & 9.9 & 9.6 & -\\
0.09\tablenotemark{a}&$10^8$  & 256 & - & 92 & 10.8 & 10 & 9.8 & -\\
\hline
0.06 & $10^8$ & - & 10 & 15 & 8.2 & 6.9 & 6.4 & 300\\
0.06 & $10^8$ & - & 50 & 15 & 8.1 & 6.4 & 5.6 & 264\\
0.06 & $10^8$ & - & 100 & 15 & 8.0 & 6.1 & 5.3 & 170\\
\label{table}
\end{tabular}
\tablenotetext{a}{Obtained using sharp $k$-space filtering}
\end{table}

\subsection{Mean reionization history}

We model the mean reionization history wtih two spatially uniform mean free paths:
that corresponding to ionized bubbles, $\lambda_{\rm b}$, and
Lyman-limit systems, $\mfp$.  Ionizing 
radiation is attenuated by their superposition,
so that $\lambda^{-1}_{\rm mfp}=\lambda^{-1}_{\rm b}+\mfp^{-1}.$  
The spatially averaged ionizing flux is $F(z)=\lambda_{\rm mfp}(z)\epsilon(z)$,
where $\epsilon(z)$ is the ionizing photon emissivity.  We
set  $\epsilon(z)=\zeta n_{H,0} \dot{f}_{\rm coll}(z)$, where $\zeta$ is
number of ionizing photons per collapsed hydrogen atom, corrected for recombinations outside of
absorption systems.  The mean ionization rate is
\begin{equation} 
\frac{d{{x}}}{dt}=\zeta\dot{f}_{\rm
  coll}(z)\frac{\mfp(z)}{\mfp(z)+\lambda_{\rm b}({x})}. 
\label{dxdt}
\end{equation}
When $\mfp\gg \lambda_{\rm b}$, photons typically reach the
edges of bubbles without being absorbed by intervening Lyman-limit
systems, and the photoionization rate is equal to the emission
rate.  When $\mfp\ll \lambda_{\rm b}$, the
probability of a photon reaching the edge of a bubble is $\mfp/\lambda_{\rm
  b}$, and the ionization rate is suppressed.  

A more realistic treatment would involve allowing $F$,
$\lambda_b$, and $\lambda_{\rm abs}$ to vary spatially, obtaining a
spatially-dependent solution of equation (\ref{dxdt}), which can then
be averaged to find the global reionization history. We have chosen
the simpler uniform model of equation (\ref{dxdt}) as a starting
point. Although this choice does not affect our results on the
morphology of the ionization field at fixed ionized fraction, the
assumptions underlying equation (\ref{dxdt}) should be kept in mind
when interpreting the reionization history and photon consumption
rates we find. 

\begin{figure}
\begin{centering}
\includegraphics[width=0.45\textwidth]{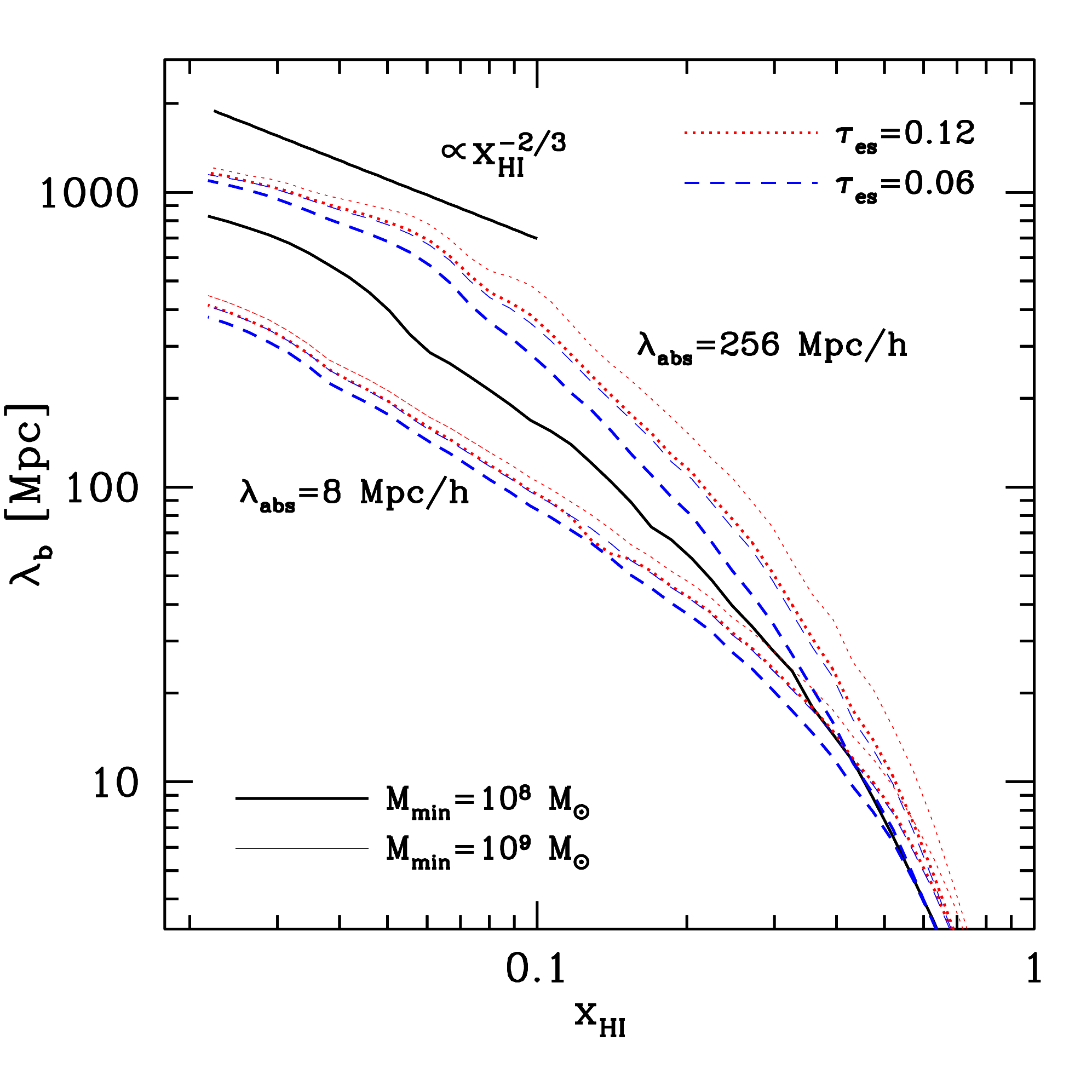}
\caption{Bubble mean free path, $\lambda_{\rm b}$, as a function of 
mean neutral fraction, $x_{\rm HI}$.  All models exhibit a 
characteristic scale of $\sim 10$~Mpc at the half-ionized 
epoch.  During percolation, lower values 
of $\lambda_{\rm abs}$ lead to lower values of $\lambda_{\rm
b}$.  
The solid line corresponds to the evaporating minihalo model with
$t_{\rm ev}=100$~Myr.  At $x_{\rm HI}<0.1$ the relation
approaches approximately $\lambda_{\rm b}\propto x_{\rm HI}^{-2/3}$.
}
\label{bubblemfp}
\end{centering}
\end{figure}
\label{sec:history}
\subsection{Spatial variations}

Our three-dimensional model is based on that of
\citet{furlanetto/etal:2004}, later extended to three dimensions
by \citet{zahn/etal:2007}.  Its main assumption is that a
region is fully ionized if its collapsed fraction
is greater than some threshold,  $\zeta f_{\rm coll} > 1$.
As shown in \citet{alvarez/etal:2009}, 
by smoothing the linear density
field over a range of scales, 
one can efficiently determine when each point 
is first reionized, $z_r$.  We do not smooth
over scales with radii larger than $\mfp$, since absorption systems
shield radiation from these distances.  We assume that the absorption
systems contribute a spatially uniform opacity, and therefore the
scale beyond which we do not smooth is the same everywhere.
\begin{figure*}
\begin{centering}
\includegraphics[width=\textwidth]{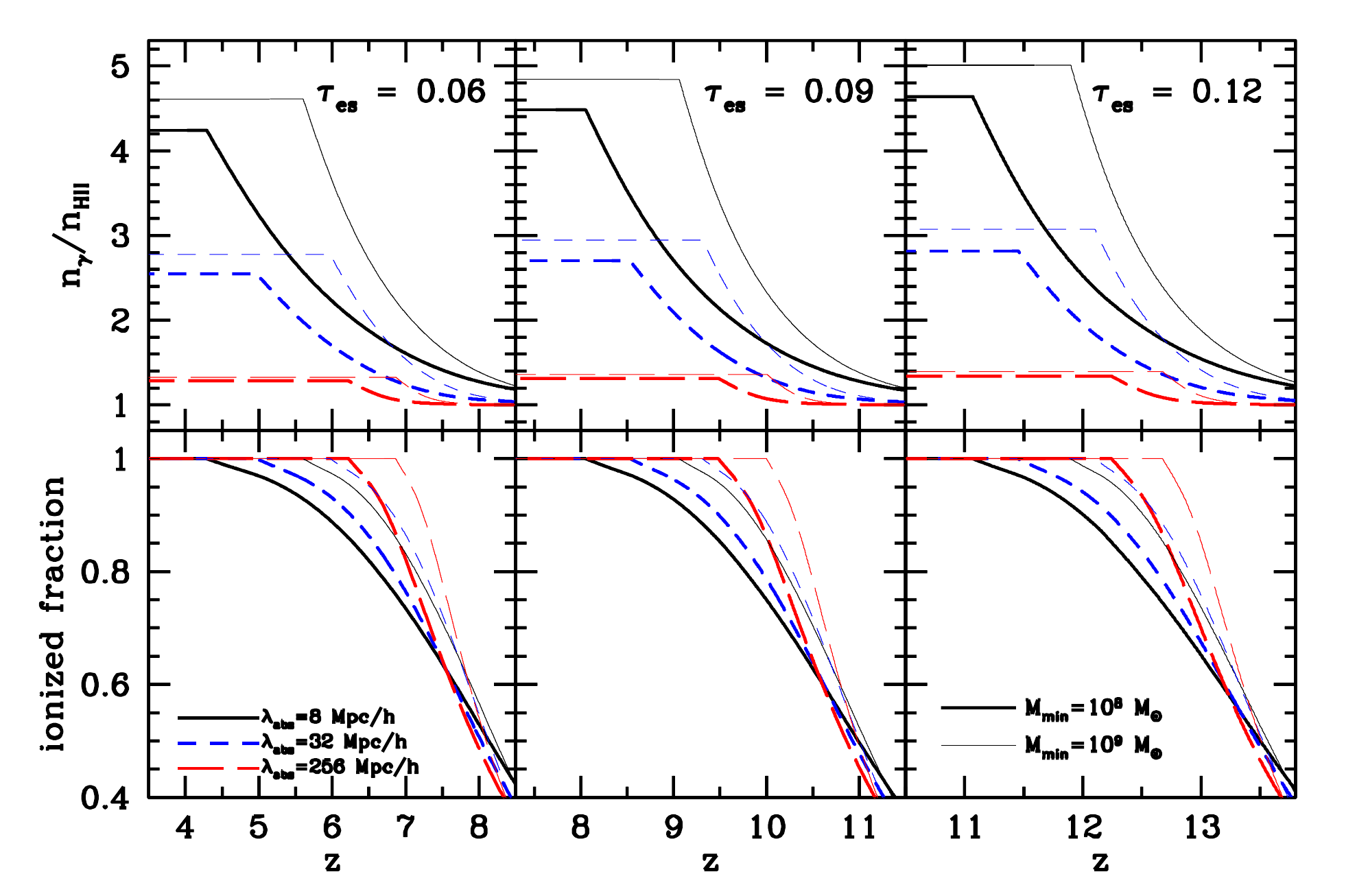}
\caption{Global reionization histories (bottom) and photon 
consumption (top). Different colors 
indicate different values of $\mfp$, while the panels indicate 
different Thomson scattering optical depths, $\tau_{\rm es} = 0.06$, 
0.09, and 0.12, from left to right.  Note that in all cases the early  
part of the reionization history is insensitive to $\mfp$, 
when the bubbles are relatively smaller.  Lower values of $\mfp$ delay 
the end of reionization. 
}
\label{history}
\end{centering}
\end{figure*}

This approach is desirable because the order in which
points are ionized is the same as that in which the
\citet{furlanetto/etal:2004} criterion is first met around each point.
However, the reionization redshifts do not generally result
in the correct average ionization fraction for a sharp smoothing
filter in real space
\citep[see below for solutions using a sharp filter in $k$-space,
which does conserve photons;][]{zahn/etal:2007}.  
To obtain a self-consistent solution, we determine a correction to the
reionization redshift at each point, $z_{r,c}(z_r)$, by matching the
simulated reionization history to the global reionization history,
\begin{equation}
\int_{z_{r}}^\infty\frac{dp}{dz}dz =
\int_{z_{r,c}}^\infty\frac{d{x}}{dt}\frac{dt}{dz}dz,
\label{match}
\end{equation}
where  $dp/dz$ is the distribution of uncorrected simulated reionization
redshifts, and $d{{x}}/dt$ is given by equation (\ref{dxdt}).  
This calibration procedure does not change the overall topology we
find at fixed ionized fraction, but does change the overall
reionization history.

We calculate $\lambda_{\rm b}({x})$ by random raytracing. If a ray
starts in an ionized region, it is extended in a random direction
until it reaches a neutral cell. The bubble mean free path is then the
average over the lengths of all such rays. The average converges with
on the order of $10^4$ rays.   

About 30 logarithmically-spaced smoothing scales
are sufficient to converge on the reionization morphology.
Because we determine the reionization redshifts 
simultaneously, the amount of data stored and the operation count are
greatly reduced compared to determining the ionization field at each
redshift.  Since the assumption that points are either highly-ionized
or neutral is made in both cases, no information is lost by following
the more efficient approach.   

\subsection{Sharp $k$-space filtering}

Using a sharp $k$-space filter leads to a method which does
conserve photons \citep{zahn/etal:2007} in the long mean free path
limit, at the expense of using an unphysical filter which ``rings'' in
real space, and an ambiguity in the assignment of a smoothing scale
$R_f$ to the filter $k_f$ -- particularly troubling given that we want
to exclude scales $R_f >\mfp$. However, for comparison we have
calculated a few cases with sharp $k$-space filtering, shown in Table
1.  

The ionization history is found by integrating a diffusion equation
for the distribution in density of points that have not crossed the
barrier \citep{bond/etal:1991}, 
\beq
\frac{\partial \Pi(\Lambda,\deltat,z)}{\partial \Lambda}=\frac{1}{2}
\frac{\partial^2\Pi}{\partial\deltat^2}-\frac{\partial
  \deltat}{\partial\Lambda}
\frac{\partial\Pi}{\partial\deltat},
\label{diffusion}
\eeq
with boundary condition $\Pi(\Lambda,0,z)=0$, where $\deltat\equiv
B(\Lambda,z)-\delta$, $\delta$ is the density contrast, 
$B(\Lambda,z)=\delta_c(z)-{\rm
  erf}^{-1}(1-\zeta^{-1})\sqrt{2(\sigma_{\rm min}^2-\Lambda)}$ is the 
bubble barrier \citep{furlanetto/etal:2004}, and $\sigma^2_{\rm min}$
is the variance of the $z=0$ linear density field at the scale of
$M_{\rm min}$, with the growth of structure encoded in the time
dependence of $\delta_c(z)$.   

We smooth only over scales larger than $\mfp$ by starting integration
of equation \ref{diffusion} at $\Lambda_i\equiv \sigma^2(\mfp)$ -- the 
variance on the scale of the mean free path is matched to the one
obtained using a top-hat filter in real space -- with initial
condition 
\beq 
\Pi(\Lambda_i,\deltat,z)=\frac{1}{\sqrt{2\pi\Lambda_i}}
\exp\left[-{\frac{(\deltat-B(\Lambda_i,z))^2}{2\Lambda_i}}\right].
\eeq 
The total ionized fraction is obtained by integrating over the
distribution at $\Lambda=\sigma_{\rm min}^2$, 
\beq
x(z)=1-\int_0^\infty \Pi(\sigma_{\rm min}^2,\deltat,z)d\deltat.  
\eeq
If the initial condition for $\Pi$ is specified at $\Lambda=0$,
(i.e. starting at a scale $\rightarrow \infty$), then this corresponds
to smoothing over all scales, and 
\beq x(z)=\zeta{\rm erfc} \left[
\delta_c(z)/\sqrt{2\sigma_{\rm min}^2}
\right]=\zeta f_{\rm
  coll}(z).  
\eeq

\subsection{Evolving $\lambda_{\rm abs}$ due to minihalos}

To model an evolving value of $\lambda_{\rm abs}$, we assume that the
absorption systems are minihalos\citep[e.g.,][]{abel/mo:1998}.  Given that the halo
cross section increases as $M^{2/3}$, while $dn/d\ln M\propto M^{-1}$,
small objects should dominate the mean free path, subject to
photoevaporation effects
\citep[e.g.,][]{haiman/etal:2001,shapiro/etal:2004}.  

Consider a uniform distribution of dark matter halos with a number
density $n_h(z)\equiv dn/d\ln M_h$.  We will assume that in neutral
regions, halos at mass $M_h$ have gas fractions of unity and would
serve as Lyman-limit absorptions systems, while in ionized regions,
such halos survive for a time $t_{\rm ev}$ before being evaporated by
the ionizing background.  The number of absorption systems evolves
according to
\beq
\frac{dn_{\rm abs}}{dz}=\frac{d\ln x}{dz}\left(n_{\rm h}-n_{\rm abs}\right)
+\xi_{\rm ev} (1+z)^{-5/2} n_{\rm abs}(z),
\label{nabs}
\eeq
where $\xi_{\rm ev}\equiv (H_0\Omega_{\rm m}^{1/2}t_{\rm
  ev})^{-1}\simeq 260\ (t_{\rm ev}/100\  {\rm Myr})^{-1}$.  We choose
$M_h=5.7\times 10^3 M_\odot[(1+z)/10]^{3/2}$,
corresponding to the cosmological Jeans mass, and consider three 
models, in which $t_{\rm   ev}=10$, 50, and 100~Myr.  These
timescales are relatively long compared with the rather short
photoevaporation times for low-mass halos found by
\citet{shapiro/etal:2004}.  However, if reionization is
``photon-starved'', as observations are suggesting
\citep[e.g.,][]{bolton/haehnelt:2007a}, then the flux could be low at
the end of reionization, leading to longer evaporation times.   

The mean free path is 
\beq
\lambda^{-1}_{\rm abs}(z)=\pi R^2_{\rm vir}(z)n_{\rm abs}(z),
\label{lambdaabs}
\eeq
where $R_{\rm vir}$ corresponds to the halo virial radius.
However, the boundaries of the absorption systems are not
well-defined due to hydrodynamic effects and departures from spherical
symmetry.   For our purposes, the minihalo model we describe here
is sufficient to incorporate the effect of an evolving absorption
system mean free path in our calculations.  More accurate modeling
of absorption systems during reionization will need to incorporate the
radiative transfer of ionizing radiation in cosmological hydrodynamics
simulations which resolve the Jeans mass. 
 
We start with an initial guess for $\lambda_{\rm abs}(z)$, only
allowing points to cross the barrier at some scale and redshift if
that scale is below the current value of $\lambda_{\rm abs}$.  The
values obtained for $\lambda_b(x)$ from the simulated ionization field
are then fed back into equations \ref{dxdt}, \ref{nabs}, and
\ref{lambdaabs}, and the process is reapeated until convergence is
reached.    

\section{Results}

Fig.~\ref{bubblemfp} shows the bubble mean free path as a function of
the neutral fraction for several parameter choices.   At the
half-ionized epoch, all models exhibit a characteristic bubble scale
of $\lambda_{\rm b}\simeq 10$~Mpc, with models with rarer sources
(i.e. increasing efficiency of ionizing radiation) leading to somewhat
larger bubbles.  During percolation, most of the variation in $\lambda_{\rm b}({x_{\rm HI}})$
comes from variation of $\mfp$, which is most pronounced at $x_{\rm
  HI}\leq 0.2$. The large values of $\lambda_b$ indicate that
simulation volumes in excess of several hundred Mpc on a side are
necessary in order to properly model the last ten per cent of the
reionization process.  At $x_{\rm HI}<0.1$, $\lambda_b\sim x_{\rm
  HI}^{-2/3}$, in agreement with the model of
\citet{miralda-escude/etal:2000}, in which a constant number of
neutral clouds decrease in size at the same fractional rate at the end
of reionization. 

The reionization history for all of the fixed $\lambda_{\rm abs}$
models is shown in Fig.~\ref{history}.  The early evolution is not
sensitive to  $\mfp$, because the photon mean free path is determined
by the size of the ionized bubbles themselves.  At later times, when the
bubble sizes become comparable to $\mfp$, the reionization history
is sensitive to $\mfp$, as can be seen by comparing the evolution for
the bracketing values of 8 Mpc/h and 256 Mpc/h.  For the
$\mfp=8$~Mpc/h case, the redshift at which the overlap occurs is
delayed by $\Delta z\sim 1.5$, relative to $\mfp=256$~Mpc/h, for
$\tau_{\rm es}=0.09$ (see also the ``$z_{\rm ov}$'' entries in Table 1). 

Also shown in Fig.~\ref{history} is the number of ionizing photons per
ionized atom,  
\beq
\frac{n_\gamma(z)}{n_{\rm HII}(z)}=\frac{\zeta f_{\rm coll}(z)}{{x}(z)}=1+\frac{1}{{x}(z)}\int_0^{{x}(z)}\frac{\lambda_{\rm b}(x')}{\mfp(x')}dx'.
\eeq
As $\mfp\rightarrow \infty$, ${x} \rightarrow \zeta f_{\rm
  coll}$ and $n_\gamma/n_{\rm HII}\rightarrow 1$.  We only count photons up
to the moment of overlap (${x}=1$), as we are concerned here with the
consumption of ionizing photons during the reionization process.  This
integral converges, since at the end of reionization we expect $\lambda_{\rm
  b}(x)\propto (1-x)^{-2/3}$ as the remaining diffuse neutral hydrogen
patches dissappear.  For $\mfp=8$~Mpc/h, about
3 photons per atom are consumed in the absorption systems by $z_{\rm ov}$.
In the longest mean free path case, $\mfp=256$~Mpc/h, the
corresponding fraction is about a half. 
 
\begin{figure}
\begin{centering}
\includegraphics[width=0.45\textwidth]{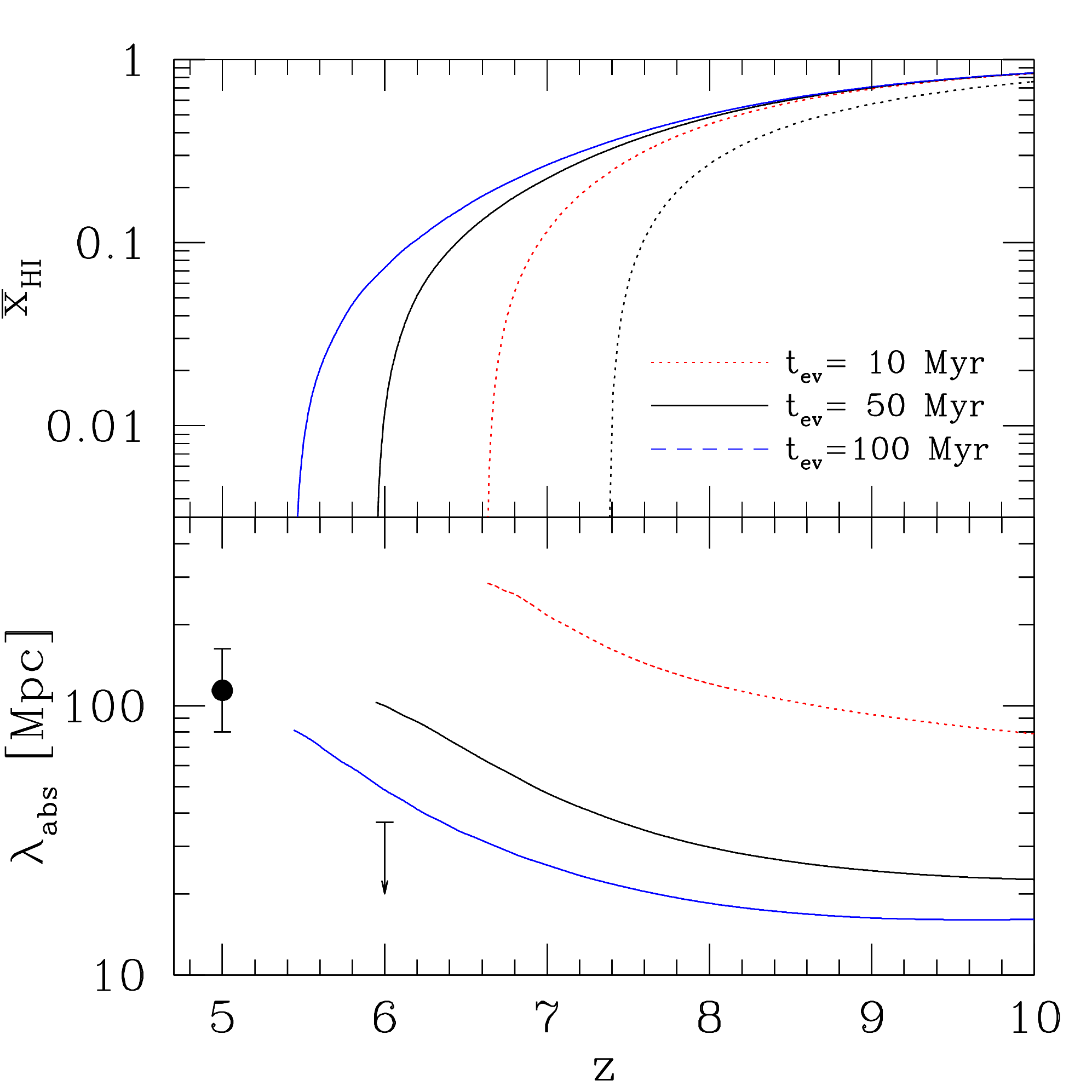}
\caption{
(top) Neutral fraction vs. redshift for the three evolving
$\lambda_{\rm abs}$ models considered here, in addition to the
reference case, for which $\lambda_{\rm abs}\rightarrow \infty$
(right-most dotted curve). (bottom) Absorption mean free path
vs. redshift.  As the last neutral patches are cleared away, so are
the last reservoirs of cold neutral gas in which the minihalos can
survive as absorption systems, so $\lambda_{\rm abs}$ increases at the
end of reionization.  As the evaporation time decreases, the
destruction rate of absorption systems increases, leading to longer
mean free paths.  Also shown are the constraints on $\mfp$ from
\citet{bolton/haehnelt:2007a} at $z=5$ and 6.
}
\label{absmfp}
\end{centering}
\end{figure}


Shown in Fig.~\ref{absmfp} is the evolution in neutral fraction and
absorption system mean free path for the three minihalo absorption
system models that we simulated.  Models with longer evaporation times
lead to a higher abundance of absorption systems, and hence shorter
mean free paths and a relative delay in the time of overlap.  The mean
free path in the $t_{\rm ev}=100$~Myr model evolves from about 40 Mpc
when $x_{\rm HI}\sim 0.1$, to about 80 Mpc at overlap.  This model is
also plotted as the solid black line in Fig.~\ref{bubblemfp}, which
shows that the instantaneous $\lambda_{\rm b}$ values for the evolving
$\lambda_{\rm abs}$ model roughly interpolate between those for the fixed
$\lambda_{\rm abs}$ models.  This can also be seen by noting that
$\lambda_0\equiv \lambda_{\rm b}(x_{\rm HI}=0.1)$ for the evolving
mean free path model, for which $\lambda_{\rm abs}(x_{\rm
  HI}=0.1)\simeq 32$~Mpc$/h$, is the same as that for the corresponding
fixed mean free path model with $\lambda_{\rm abs}=32$~Mpc$/h$,
$\lambda_0=170$~Mpc. 

\begin{figure*}
\begin{centering}
\includegraphics[width=0.47\textwidth]{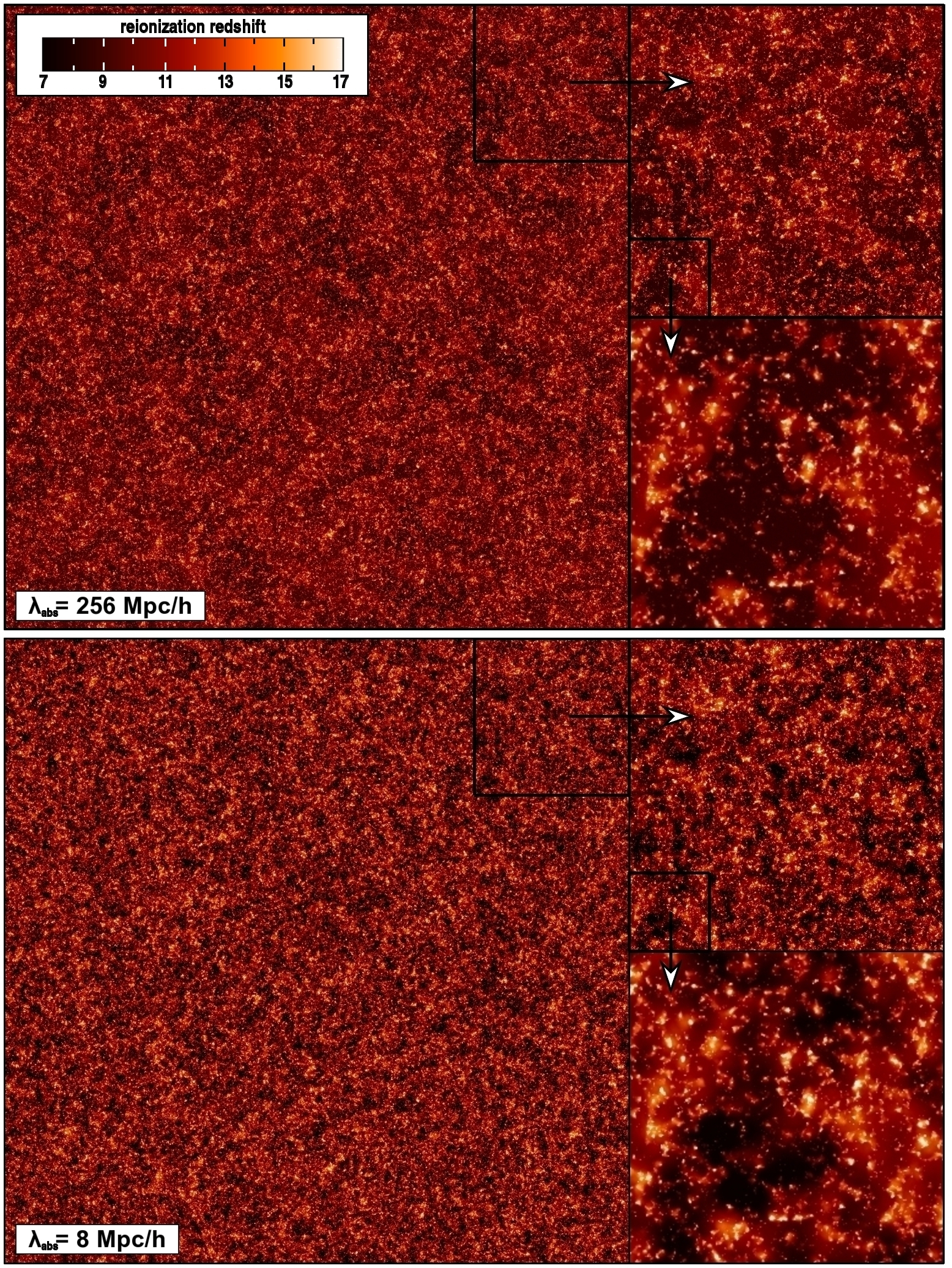}
\includegraphics[width=0.47\textwidth]{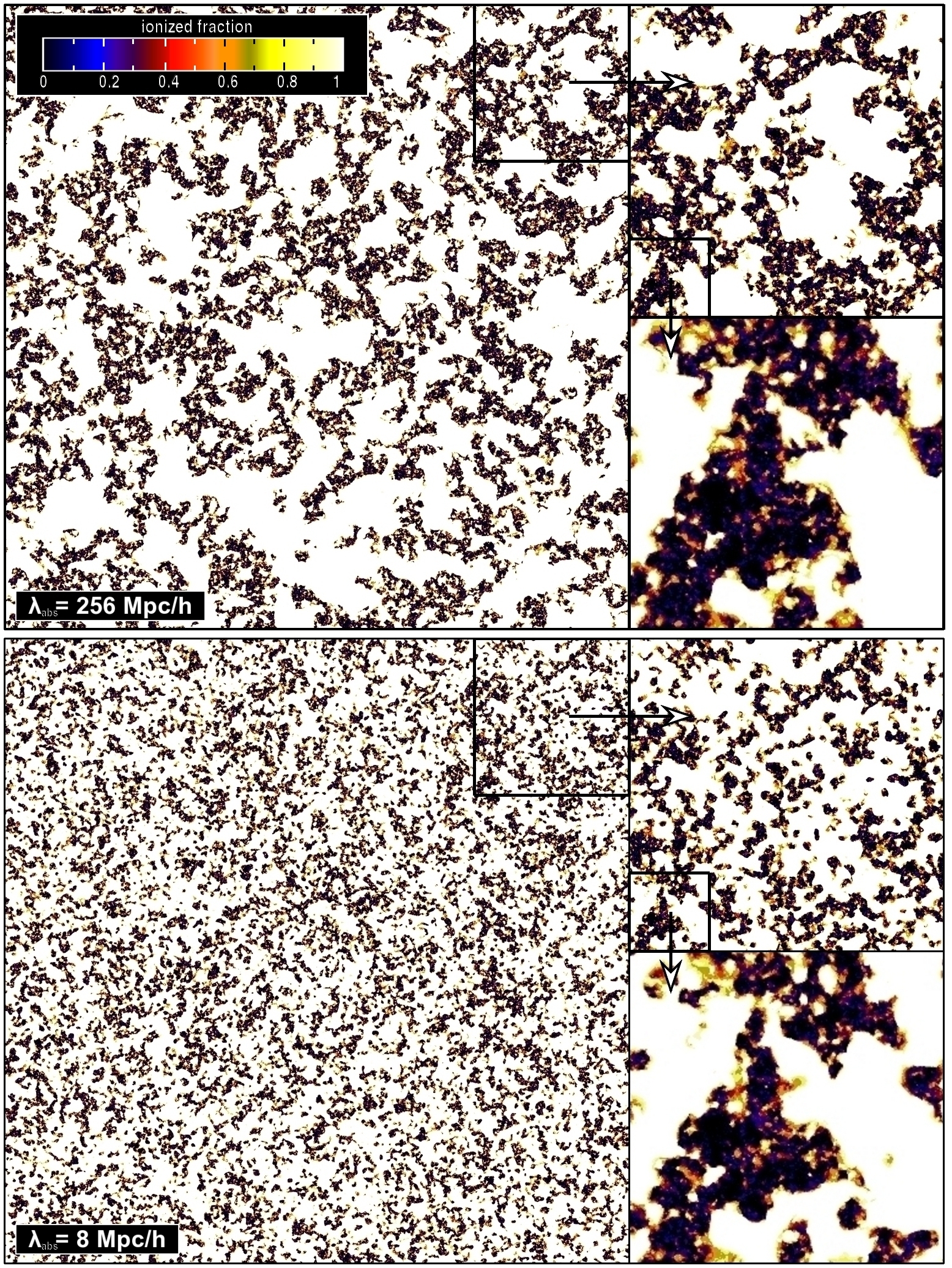}
\caption{{\em left:} Reionization redshift in a 0.5 Mpc$/h$-thick
slice for the $\mfp=256$ Mpc$/h$ simulation (top) and the $\mfp=8$
Mpc$/h$ simulation (bottom). Both cases were for $\tau_{\rm es}=0.09$
and $M_{\rm min}=10^8 M_\odot$. Clockwise beginning from top-left,
regions shown are 2 Gpc$/h$, 500 Mpc$/h$, and 125 Mpc$/h$ across.
The radii of the circles correspond to $\mfp$.
{\em right:} Same as left panels but for the projected ionized fraction 
in a 5 Mpc$/h$-thick slice at a time when the mean volume-weighted ionized
fraction is 0.75.  Note that the ionization field contains
considerably more small scale structure in the $\mfp=8$ Mpc$/h$ case.}
\vspace{1.0cm}  
\label{panels}
\end{centering}
\end{figure*}

Fig.~\ref{panels} shows the ionization field in a 5-Mpc$/h$ slice at
${x}=0.75$ for $M_{\rm min}=10^8 M_\odot$, with
$\mfp=256$~Mpc$/h$ (top) and $\mfp=8$~Mpc$/h$ (bottom).  The two cases
are quite different, with many more small neutral
patches for $\mfp=8$~Mpc$/h$ relative to $\mfp=256$~Mpc$/h$. Also
shown in Fig.~\ref{panels} is the reioniztion redshift in a
0.5-Mpc$/h$ slice.  Lower values of $\mfp$
result in a more extended reionization overlap period -- i.e. the
maxima in reionization redshift are higher, and the minima are lower.      

\section{Discussion}

We have carried out large-scale simulations of reionization in a 2
Gpc$/h$ volume, including a finite mean free path to absorption
systems.  Absorption systems have a significant 
effect on the characteristic scales at the end of reionization.  
For the Thomson scattering optical depth reported by {\em WMAP},
$\tau_{es}\simeq 0.09$, we find that the characteristic bubble size 
when the universe is 90 per cent ionized is quite sensitive to $\mfp$, 
with $\lambda_0\sim 440$~Mpc for $\mfp=256$~Mpc$/h$ and $M_{\rm
  min}=10^9 M_\odot$, while $\lambda_0\sim 94$~Mpc for
$\mfp=8$~Mpc$/h$ and $M_{\rm min}=10^8 M_\odot$.   

Calculations using a sharp $k$-space filter lead to a
more modest extension of the percolation phase (see Table 1) compared
to the reionization history obtained by the solution of equation
(\ref{dxdt}), on which the reionization histories shown in Figure 1
are based. For $\tau_{es}\simeq 0.09$, the difference in the redshift
when the universe is 90 per cent ionized between the $\mfp=8$~Mpc$/h$
and $\mfp=256$~Mpc$/h$ cases is only $\Delta z_{0.9}\sim 0.3$, while the
corresponding case using a sharp real space filter and equation
(\ref{dxdt}) is $\Delta z_{0.9}\sim 0.7$. The delay in the overlap time in
both cases is $\Delta z_{\rm ov}\sim 1.5$. 

Using either sharp $k$-space smoothing or equation (1) to model the
effect of absorption systems on the reionization history has
drawbacks. While equation (1) conserves photons in the long mean free
path limit, a mean flux and opacity are assumed, and further work will
be necessary to validate and/or improve upon the approach layed out in
section \ref{sec:history}. Using a sharp $k$-space filter results in a reionization history
which also conserves photons in the long mean free path limit
($k_F\longrightarrow 0$), but at the expense of using an oscillatory
filter which can ``leak'' photons from high density regions into lower
density ones, and for which there is no unique choice for the relation
between $\mfp$ and $k_F$. While our results for the timing and photon
consumption of the end of reionizaiton are obtained from equations (1)
and (9), it is important to note that more work will be necessary to
test their accuracy. However, results for the morphology of
reionization (Figures 1 and 4) do not depend on the model for the
global history, and are therefore more robust. 

Our results are consistent with those of  \citet{furlanetto/oh:2005},
in which consumption of ionizing photons  in dense systems extends the
end of reionization considerably.  \citet{choudhury/etal:2009} used
the semi-numerical approach in a volume~$100$~Mpc$/h$ across, and
found a similar trend with decreasing $\mfp$, indicating a transition
to an ``outside-in'' morphology near the end of reionization.  We find
that neutral patches may also remain in voids, where formation of
ionizing sources is delayed and radiation from the nearest sources is
shielded.   Previous studies that modeled the physical origin of
absorption systems as minihalos
\citep[e.g.,][]{ciardi/etal:2006,mcquinn/etal:2007} also found similar
effects to those we find here, although these studies were more
focused on the intermediate stages of reionization and on smaller
simulated volumes, where the photon mean free path was not as small
relative to the bubble sizes as we find in our low mean free path
cases in the final, percolation phase.  

As mentioned in the introduction, our approach is complementary to
that taken by \citet{crociani/etal:2011}, in which the semi-numerical
approach was used to determine the distribution of absorbers during
reionization. They found that their spatial distribution is quite
inhomogenous, owing both to intrinsic density fluctuations in the IGM,
as well as fluctuations in the ionizing radition field -- regions far
from sources have a relatively low flux, resulting in a higher
abundance of absorbers. This affect was also pointed out by
\citet{mcquinn/etal:2011}, who used high-resolution hydrodynamic
simulations post-processed with radiative transfer to determine the
mean free path as a function of flux, finding a strongly nonlinear
dependence of the mean free path on the emmissivity of ionizing
radiation. 

These effects imply an important improvement to our model is not only
a time-varying mean free path, but also a spatially varying
one. To accomplish this, more work will need to be done on the
``sub-grid'' physics during reionization, using the results of
high-resolution cosmological simulations with radiative transfer of a
background radiation field, coupled to the hydrodynamics of the gas.
However, given that the final patches to be ionized are those most
distant from the most luminous sources, the delay in the very end of
reionization that we find here is likely to persist when the
inhomogeneity of the IGM opacity due to absorption systems is properly
taken into account. 

The lingering neutral clouds we find would further complicate
interpretation of quasar absorption spectra at $z\sim 6$
\citep[e.g.,][]{bolton/haehnelt:2007b, lidz/etal:2007,
  alvarez/abel:2007, wyithe/etal:2008, furlanetto/mesinger:2009}. As
discussed  by \citet{mesinger:2009}, gaps in quasar spectra can come
from either these neutral clouds or mostly ionized but optically-thick
absorption systems.  The line-of-sight abundance of these neutral
clouds at fixed ionized fraction increases with decreasing absorption
system mean free path. 

As shown by \citet{weinmann/etal:2007} and \citet{alvarez/etal:2009},
the Local Group may have been reionized by external sources, i.e. the
progenitors of the Virgo Cluster at a distance of $\sim$~20 Mpc.  If
$\mfp<20$~Mpc, however, local reionization would have
been delayed until Local Group prongenitors
had formed in sufficient number \citep[e.g.,][]{munoz/etal:2009}.
Since the satellite population is sensitive to the timing of
reionization \citep{busha/etal:2010}, lower values of $\mfp$ could
imply a higher local satellite abundance.  

Finally, our results have important implications for the
interpretation of temperature fluctuations in the cosmic microwave
background at multipoles $l\sim 3000$ \citep{das/etal:2011,
  reichardt/etal:2011}, through the kinetic Sunyaev-Zel'dovich (kSZ) effect 
\citep[e.g.,][]{mcquinn/etal:2005, iliev/etal:2007}.
In particular, \citet{mesinger/etal:2011b} used nearly the same three
parameters and approach as we used here for their parameter study of
the dependence of the kSZ signal on the reionization
scenario. Reducing the mean free path shifts power to higher
multipoles because of the accompanying decrease in the bubble size
that we find. Although the kSZ effect is only just beginning to
constrain the duration of reionization \citep{zahn/etal:2011b}, more
detailed future observations and analysis will begin to provide
constraints on the patchiness in addition, and understanding the
effect of absorption systems on reionization will be crucial.

\acknowledgments{We thank M.~Haehnelt, M.~McQuinn,
  A.~Mesinger, R.~Thomas, G.~Vasil for useful discussions, and
  J.~Chluba for providing the diffusion solver from CosmoRec. The
  simulations in this paper  
  were performed on the GPC supercomputer at the SciNet HPC
  Consortium.  SciNet is funded by: the Canada Foundation for
  Innovation under the auspices of Compute Canada; the Government of
  Ontario; Ontario Research Fund - Research Excellence; and the
  University of Toronto. This work was partially supported by NASA
  ATFP grant NNX08AH26G, NSF AST-0807312, and NSF AST-0808398.}

\end{document}